\documentclass[12pt]{article}
\usepackage{amssymb}
\usepackage{epsfig}

\catcode`\@=11
\def\numberbysection{\@addtoreset{equation}{section}
        \def\theequation{\thesection.\arabic{equation}}}

\def\beq{\begin{equation}}
\def\eeq{\end{equation}}
\numberbysection
\begin{document}
\begin{titlepage}
\begin{center}
\hfill  \\
\vskip 1.in {\Large \bf Euclidean black holes and spin connection} \vskip 0.5in P. Valtancoli
\\[.2in]
{\em Dipartimento di Fisica, Polo Scientifico Universit\'a di Firenze \\
and INFN, Sezione di Firenze (Italy)\\
Via G. Sansone 1, 50019 Sesto Fiorentino, Italy\\
Email address: valtancoli@fi.infn.it}
\end{center}
\vskip .5in
\begin{abstract}
The Euclidean method is usually discussed in the context of the metric avoiding the typical delta of the conical singularity. We introduce a new way to calculate the Hawking temperature using the vierbein and spin connection.
The conical singularity is seen globally through an effect on parallel transport, the so called holonomy of the spin connection. The period of the Euclidean time is calculated requiring that the holonomy of the spin connection
is trivial at the event horizon.
\end{abstract}
\medskip
\end{titlepage}
\pagenumbering{arabic}
\section{Introduction}

The complex rotation of the time coordinate to imaginary values gives many insights into the quantum nature of space-time. In a generic curved space-time there is no preferred time coordinate and the Wick rotation does not make
much sense. Only in the special case of static or stationary space-times such as the Schwarzschild metric or the Kerr metric equipped with Killing vectors it is possible to rigorously define the Wick rotation.

Indeed the analytic continuation of the time coordinate to imaginary values $ t \rightarrow it $ is an effective method to discuss the close link between black holes and thermodynamics. The Euclidean method produces results in agreement with those obtained by completely different techniques and correctly reproduces both the Hawking temperature and the entropy of black holes at the
semiclassical level \cite{1}-\cite{2}-\cite{3}-\cite{4}-\cite{5}. Among these alternative techniques we can cite 't Hooft's brick wall method \cite{6}, according to which the entropy of a black hole can be linked to a thermal gas of quantum field theory excitations which propagate outside the horizon. 't Hooft introduced the so-called brick wall, a fixed boundary near the horizon within which quantum fields do not propagate. In \cite{7} it has been shown that Pauli Villars' regularization automatically implements a cutoff in the 't Hooft computation making the introduction of the brick wall unnecessary.

But the natural way to formulate the thermodynamics of black holes is to use the Euclidean path integral approach. The Wick rotation of the time coordinate is usually discussed within the metric formalism. Avoiding the conical singularity that is naturally formed by sending $ t \rightarrow it $, leads to the periodicity of Euclidean time with a period linked to the inverse of the Hawking temperature. Gibbons and Hawking \cite{1}, using this periodicity of Euclidean time, computed the path integral of the action of the classical Euclidean gravity rediscovering the correct expression of the entropy of the black hole. Unfortunately this method is difficult to implement for black holes with many degrees of freedom.

The purpose of this article is to explore Wick rotation using the vierbein and spin connection formalism. At the vierbein level we achieve the same aim with a global measurement of the conical singularity by parallel transport, i.e. calculating the holonomy of the spin connection. In this case to avoid the conical singularity we have to require that the Wick rotated holonomy of the spin connection is trivial at the event horizon ( see also \cite{8} ).
This condition reproduces the correct Hawking temperature of the black hole and, as we shall see, is in general faster to apply than the metric formalism. The choice of the path to compute the holonomy is practically dictated by which component $\mu$ of the spin connection $\omega_{\mu}^{ab}$ survives at the horizon.

The article is organized as follows. In section $ 2 $ we first discuss the $ BTZ $ black hole in $ 2 + 1 $ dimensions of which we give a detailed analysis of how to derive the Hawking temperature from the spin connection. We find that in addition to a Euclidean time period there is an extra period of the angle coordinate $ \phi $. The calculation can be done in two different ways but in the end the same result is obtained.

In section $ 3 $ we extend these results to the black hole with angular momentum in $ 3 + 1 $ dimensions, the Kerr metric, in which we obtain results analogous to the simplest case in
 $ 2 + 1 $ dimensions. In this case the method is faster than working directly within the metric formalism.

\section{BTZ black hole}

The Euclidean method is usually discussed within the metric formalism avoiding the conical singularity that appears in the analytic continuation of the time coordinate to imaginary values $ t \rightarrow it $.
The aim of our article is to demonstrate that the Wick rotation of black hole metrics can be understood using the first order formalism ( vierbein and spin connection ). This method is particularly useful in the case of black holes with angular momentum. Let us first consider the $BTZ$ black hole whose metric is defined by:

\begin{eqnarray} ds^2  & = & - N^2 \ dt^2 \ + \ N^{-2} \ dr^2 \ + \ r^2 \ ( \ d\phi \ + \ N^\phi \ dt \ )^2 \nonumber \\
N^2 & = & - M \ + \ \frac{r^2}{l^2} \ + \ \frac{J^2}{4r^2} \ \ \ \ N^\phi \ = \ - \ \frac{J}{2 r^2} \nonumber \\
M & = & \frac{r^2_{+} \ + \ r^2_{-}}{l^2} \ \ \ \ \ J \ = \ \frac{2 r_{+} r_{-}}{l} .
\label{21} \end{eqnarray}

The $BTZ$ black hole contains two special rays $ r_{+} $ and $ r_{-} $, the outer one $ r_{+} $ is the event horizon while $ r_{-} $ is the inner horizon , analogous to an ergosphere.

The corresponding vierbein is defined by:

\begin{eqnarray}
\sigma^0 & = & N(r) \ dt \nonumber \\
\sigma^1 & = & \frac{dr}{N(r)} \nonumber \\
\sigma^2 & = & r d\phi \ - \ \frac{J}{2r} \ dt .
\label{22} \end{eqnarray}

The spin connection $ \omega_\mu^{ab} $ can be expressed as a linear combination of the components of the vierbein ( eq. ( \ref{22} )):
\begin{eqnarray}
\omega^{01} & = & \frac{\partial}{\partial r} N(r) \ \sigma^0 \ - \ \frac{J}{2 r^2} \ \sigma^2 \nonumber \\
\omega^{12} & = & - \left[ \ \frac{J}{2r^2} \sigma^0 \ + \ \frac{N(r)}{r} \ \sigma^2 \ \right] \nonumber \\
\omega^{02} & = & - \frac{J}{2r^2} \ \sigma^1 . \label{23}
\end{eqnarray}

As we can see, most of the coefficients in front of the components of the vierbein are either constant or tend to zero in the limit $ r \rightarrow r_{+} $. Only $ \omega^{01} $ contains a coefficient which becomes singular in the limit $ r \rightarrow r_{+} $.

 Wick rotation in the case of black hole metrics with angular momentum is more complicated than usual, because in addition to transform the time coordinate it is also necessary to modify the angular momentum

\begin{eqnarray} t & \rightarrow & it_E \nonumber \\
J & \rightarrow &   - i J_E \ \ \ \ r_{-} \rightarrow - i r_{-}
\label{24} \end{eqnarray}

to keep the metric real:

\begin{eqnarray} ds^2  & = & N^2_E \ dt_E^2 \ + \ \frac{dr^2}{N^2_E} \ + \ r^2 \ ( \ d\phi \ - \frac{J_E}{2 r^2} \ dt_E \ )^2 \nonumber \\
N^2_E & = & \frac{( r^2 - r^2_{+} )( r^2 + r^2_{-} ) }{ l^2 r^2} \ = \  - M \ + \ \frac{r^2}{l^2} \ - \ \frac{J^2_E}{4r^2} .
\label{25} \end{eqnarray}

The corresponding vierbein is defined by

\begin{eqnarray}
\sigma^0_E & = & N_E(r) \ dt_E \nonumber \\
\sigma^1_E & = & \frac{dr}{N_E(r)} \nonumber \\
\sigma^2_E & = & r d\phi \ - \ \frac{J_E}{2r} \ dt_E
\label{26} \end{eqnarray}

and the spin connection has some sign of difference from the Minkowskian case ( eq. ( \ref{23} )):

\begin{eqnarray}
\omega^{01}_E & = & \frac{\partial}{\partial r} N_E(r) \ \sigma^0_E \ + \ \frac{J_E}{2 r^2} \ \sigma^2_E \nonumber \\
\omega^{12}_E & = & - \left[ \ \frac{J_E}{2r^2} \sigma^0_E \ + \ \frac{N_E(r)}{r} \ \sigma^2_E \ \right] \nonumber \\
\omega^{02}_E & = & \frac{J_E}{2r^2} \ \sigma^1_E . \label{27}
\end{eqnarray}

In the Euclidean case we obtain that the time coordinate is periodic. In the case of the metric formalism this is obtained by avoiding the conical singularity, but in the case of the $BTZ$ black hole this requires some complicated manipulations on the metric.  At the level of the vierbein we realize this with a global measure of the conical singularity by parallel transport, i.e. the holonomy of the spin connection. The period is obtained by requiring that the holonomy of the spin connection be trivial on the event horizon
$ r = r_{+} $

\beq \exp \left[ \int_\gamma \ \omega_E^{ab}|_{r = r_{+}} \ \right] \ = \ e^{2\pi i} \ = \ 1 . \label{28} \eeq

The spin connection is an antisymmetric matrix in the Lorentz indices ($ab$), therefore it is diagonalizable with all imaginary eigenvalues. If we have only one non-zero contribution

\beq \omega^{01} \ = \ - \omega^{10} \ = \ \lambda_1 \label{271} \eeq

then what physically matters are the eigenvalues

\beq \omega^{ab} \ = \ U^{-1} \left( \begin{array}{ccc} i \lambda_1 &  &  \\ & - i \lambda_1 & \\ & & 0 \end{array} \right) U , \label{272} \eeq

where $ U $ is a matrix with constant coefficients.

If we have two non-zero contributions,

\begin{eqnarray} \omega^{01} \ = \ - \omega^{10} \ = \ \lambda_1 \nonumber \\
 \omega^{12} \ = \ - \omega^{21} \ = \ \lambda_2 \label{273} \end{eqnarray}

then what matters are the eigenvalues

\beq \omega^{ab} \ = \ U^{-1} \left( \begin{array}{ccc} i \sqrt{\lambda_1^2 + \lambda_2^2} &  &  \\ & - i \sqrt{\lambda_1^2 + \lambda_2^2} & \\ & & 0 \end{array} \right) U . \label{274} \eeq

The choice of the path $ \gamma $ is strongly suggested by which component $ \mu $ of $ \omega_\mu^{ab} $ survives in the limit $ r \rightarrow r_ {+} $. Furthermore, we require that the path $ \gamma $ chosen to compute the holonomy satisfies $ \int_\gamma \ \sigma^i \ \rightarrow \ 0 $ for $ r \rightarrow r_{+} $. This implies that in particular $ \sigma^2_E \ |_{r=r_{+}} \ = \ 0 $ and therefore, in addition to varying the Euclidean time, the angle $ \phi $ must also be varied.

In this case the only contributing component of the spin connection is $ \omega^{01} $, because the factor $ (\frac {\partial} {\partial_r} N(r)) $ is singular in the
limit $ r \rightarrow r_{+} $ and overall you get a finite contribution:

\beq N_E(r) \left( \frac{\partial}{\partial_r} N_E(r) \right)|_{r = r_{+}} \ \Delta t_E \ = \ 2 \pi .
\label{29} \eeq

We obtain that Euclidean time is periodic with period $ \Delta t_E $ linked to the inverse of the Hawking temperature. The temperature formula is the one compatible with Wick rotation
$ r_{-} \rightarrow -i r_{-} $:

\beq \Delta t_E \ = \ \frac{2\pi l^2 r_{+}}{r^2_{+}+ r^2_{-}} \ \ \ \ \ \ \  T_H \ = \ \frac{r^2_{+}+ r^2_{-}}{2\pi l^2 r_{+}} .
\label{210} \eeq

Furthermore from the condition that $ \sigma^2_E \ |_{r=r_{+}} \ = \ 0 $

\beq \sigma^2_E \ |_{r=r_{+}} \ = \ 0 \ \ \ \ \rightarrow \ \ \ \ \Delta \phi \ = \ \frac{J_E}{2 r^2_{+}} \ \Delta t_E \ = \ \frac{2 \pi l r_{-}}{r^2_{+}+ r^2_{-}} \label{211} \eeq

we obtain that the variation of the coordinate $ \phi $ necessary to obtain a trivial holonomy is linked to the factor $ r_{-} $ and therefore if the angular momentum is zero this too disappears.

Alternatively, the problem can be solved by diagonalizing the Euclidean metric ( eq. ( \ref{25} )):

\begin{eqnarray}
t'_E & = & r_{+} \ t_E \ + \ l \ r_{-} \ \phi \nonumber \\
\phi' & = & r_{+} \ \phi \ - \ r_{-} \ \frac{t_E}{l} , \label{212}
\end{eqnarray}

from which we obtain

\beq g_{00} \ = \ \frac{r^2 - r^2_{+}}{l^2 ( r^2_{+} + r^2_{-} ) } \ \ \ \ g_{02} \ = \ 0 \ \ \ \ g_{22} \ = \ \frac{r^2 + r^2_{-}}{( r^2_{+} + r^2_{-} )} .
\label{213} \eeq

In this case the vierbein takes the following form

\begin{eqnarray}
\sigma^0_E & = & \frac{1}{( r^2_{+} + r^2_{-} )^\frac{3}{2}} \ \left\{ \left[ \ r^2_{+} \ N_E(r) \ + \ r^2_{-} \ \left( \frac{r^2-r^2_{+}}{ l r }
\right) \right] dt'_E \ + \ l \ r_{+} r_{-} \ \left[ \frac{r^2 + r^2_{-} }{ l r } \ - \ N_E(r) \ \right] d \phi' \ \right\} \nonumber \\
\sigma^1_E & = & \frac{dr}{N_E(r)} \nonumber \\
\sigma^2_E & = & \frac{1}{( r^2_{+} + r^2_{-} )^\frac{3}{2}} \ \left\{ r_{+} r_{-} \ \left[ \frac{r^2 - r^2_{+} }{ l r } \ - \ N_E(r) \ \right] d t'_E \ + \ l \ \left[ \ r^2_{-} \ N_E(r) \ + \ r^2_{+} \ \left( \frac{r^2+r^2_{-}}{ l r }
\right) \right] d\phi' \ \right\} \nonumber \\
& \ &
\label{214} \end{eqnarray}

and the corresponding spin connection is

\begin{eqnarray}
\omega^{01}_E & = & \frac{1}{( r^2_{+} + r^2_{-} )^\frac{3}{2}} \ \left\{ \left[ \ \frac{r^2_{+} ( r^2 + r^2_{-} )}{l^2 r} \ + \ \frac{r^2_{-}}{l} \ N_E(r) \ \right] dt'_E \ - \ r_{+} r_{-} \left[ \frac{( r^2 - r^2_{+})}{l r} \ - \ N_E(r) \right] d \phi' \ \right\} \nonumber \\
\omega^{02}_E & = & \frac{r_{+} r_{-}}{l r^2 N_E(r)} \ dr \nonumber \\
\omega^{12}_E & = & \frac{1}{( r^2_{+} + r^2_{-} )^\frac{3}{2}} \ \left\{ \ r_{+} r_{-} \left[ \frac{( r^2 + r^2_{-})}{l^2 r} \ - \ \frac{N_E(r)}{l} \ \right] dt'_E .
\ - \ \left[ \ \frac{r^2_{-} ( r^2 - r^2_{+} )}{l r} \ + \ r^2_{+} \ N_E(r) \ \right] d \phi' \ \right\} . \nonumber \\
 & \ & \label{215} \end{eqnarray}

If we evaluate the spin connection for $ r = r_{+} $ we have two non-zero contributions:

\begin{eqnarray} \omega_E^{01}|_{r_{+}} & = & \frac{1}{\sqrt{r^2_{+}+r^2_{-}}} \ \frac{r_{+}}{l^2} \ dt'_E \nonumber \\
\omega_E^{02}|_{r_{+}} & = & 0 \nonumber \\
\omega_E^{12}|_{r_{+}} & = & \frac{1}{\sqrt{r^2_{+}+r^2_{-}}} \ \frac{r_{-}}{l^2} \ dt'_E  . \label{216}
\end{eqnarray}

The eigenvalues of $ \omega^{ab}_E $ are ( applying eq. (\ref{274}) )

\beq \omega^{ab}_E
 \ = \ U^{-1} \left( \begin{array}{ccc} \frac{i}{l^2} &  &  \\ & - \frac{i}{l^2} & \\ & & 0 \end{array} \right) U . \label{217} \eeq

From the triviality condition of the holonomy of the spin connection we obtain the following system:

\begin{eqnarray}
\Delta t'_E & = & 2 \pi l^2 \nonumber \\
\Delta \phi' & = & 0 . \label{218} \end{eqnarray}

This system of periods is particularly advantageous because the effect of angular momentum is minimized since the period is purely temporal. The diagonalization of the metric has decoupled the time variable $ t'_E $ from the angular variable $ \phi' $. Going back to the old variables ( eq. (\ref{212}) ) we get a system that gives as a result

\beq \Delta t_E \ = \ \frac{2 \pi l^2 r_{+}}{r^2_{+}+r^2_{-}} \ \ \ \ \Delta \phi \ = \ \frac{2 \pi l r_{-}}{r^2_{+}+r^2_{-}} , \label{219} \eeq

the same calculated before. Thus we have a verification that the spin connection holonomy formula gives consistent results in various coordinate systems.

\section{Kerr metric}

The case of the Kerr metric has been briefly analyzed in \cite{1}. A detailed study of the conical singularity within the metric formalism can be found in \cite{4}. We will now analyze this case in the context of the spin connection holonomy which is a faster method. The Kerr metric in the Boyer-Lindquist coordinates is defined by

\begin{eqnarray} ds^2 & = & - \frac{\Delta}{\rho^2} \ ( dt - a \sin^2 \theta d \phi )^2 \ + \ \frac{ \sin^2 \theta}{\rho^2} \ (( r^2 + a^2 ) d \phi \ - \ a dt )^2 + \nonumber \\
 & + & \frac{\rho^2}{\Delta} dr^2 \ + \ \rho^2 d \theta^2  \ \ \ \ \ \ a \ = \ \frac{J}{M}\nonumber \\
 \Delta & = & r^2 - 2 M r + a^2 \nonumber \\
 \rho^2 & = & r^2 + a^2 cos^2 \theta .
  \label{31} \end{eqnarray}

Solving the equation $ \Delta \ = \ 0 $ we obtain the external and internal horizons ($ r_\pm \ = \ M \pm \sqrt{M^2 - a^2} $). Solving the equation $ g_{tt} \ = \ 0 $ we obtain the rays in which the temporal component changes sign ($ r_\pm^E \ = \ M \pm \sqrt{M^2 - a^2 \cos^2 \theta} $).

The associated vierbein is

\begin{eqnarray}
\sigma^0 & = & \frac{\sqrt{\Delta}}{\rho} \ ( dt - a \sin^2 \theta d \phi ) \nonumber \\
\sigma^1 & = & \frac{\rho}{\sqrt{\Delta}} \ dr \nonumber \\
\sigma^2 & = & \rho d \theta \nonumber \\
\sigma^3 & = & \frac{ \sin \theta}{\rho} \ (( r^2 + a^2 ) d \phi \ - \ a dt ) \label{32} \end{eqnarray}

and the spin connection

\begin{eqnarray}
\omega^{01} & = & - \frac{1}{\rho^3} \ \left[ \ \frac{- 2 M r^2 \ + \ a^2 \ [ \ M + r + ( M-r ) \cos 2 \theta \ ] }{2 \sqrt{\Delta}} \
\sigma^0 \ + \ ra \sin \theta \sigma^3 \ \right]\nonumber \\
\omega^{02} & = & - \frac{a \cos \theta}{\rho^3} \ [ \ a \sin \theta \sigma^0 \ + \ \sqrt{\Delta} \ \sigma^3 \ ] \nonumber \\
\omega^{03} & = & - \frac{a}{\rho^3} \ [ \ r \sin \theta \sigma^1 \ - \ \sqrt{\Delta} \cos \theta \sigma^2 \ ] \nonumber \\
\omega^{12} & = & - \frac{1}{\rho^3} \ [ \ a^2 \sin \theta \cos \theta \sigma^1 \ + \ r \sqrt{\Delta} \sigma^2 \ ] \nonumber \\
\omega^{13} & = & - \frac{r}{\rho^3} \ [ \ a \sin \theta \sigma^0 \ + \ \sqrt{\Delta} \sigma^3 \ ] \nonumber \\
\omega^{23} & = &  - \frac{\cot \theta}{\rho^3} \ [ \ a \sqrt{\Delta} \sin \theta \sigma^0 \ + \ ( r^2 + a^2 ) \sigma^3 \ ] .
\label{33} \end{eqnarray}

Also in this case only $ \omega^{01} $ contributes to the holonomy because the coefficient in front of $ \sigma^0 $ is proportional to ( $ \frac{1}{\sqrt{\Delta}} $ ) and therefore singular in the limit $ r \rightarrow r_{+} $,
while the other coefficients are all regular.

To perform the Wick rotation we must again transform both time and angular momentum:

\begin{eqnarray} ds^2_E & = &  \frac{\Delta}{\rho^2} \ ( \ dt_E + a \sin^2 \theta d \phi \ )^2 \ + \ \frac{ \sin^2 \theta}{\rho^2} \ ( \ ( r^2 - a^2 ) d \phi \ - \ a dt_E \ )^2 + \nonumber \\  & + & \frac{\rho^2}{\Delta} dr^2 \ + \ \rho^2 d \theta^2  \ \ \ \ \ \ t \rightarrow i t_E \ \ \ \ J \rightarrow  - i J_E \ \ \ \ a \rightarrow -i a \nonumber \\
 \Delta & = & r^2 - 2 M r - a^2 \nonumber \\
 \rho^2 & = & r^2 - a^2 cos^2 \theta ,
  \label{34} \end{eqnarray}

from which the Euclidean vierbein is

 \begin{eqnarray}
\sigma
^0_E & = & \frac{\sqrt{\Delta}}{\rho} \ ( dt_E + a \sin^2 \theta d \phi ) \nonumber \\
\sigma^1_E & = & \frac{\rho}{\sqrt{\Delta}} \ dr \nonumber \\
\sigma^2_E & = & \rho d \theta \nonumber \\
\sigma^3_E & = & \frac{ \sin \theta}{\rho} \ (( r^2 - a^2 ) d \phi \ - \ a dt_E ) . \label{35} \end{eqnarray}

The Euclidean spin connection now reads:

\begin{eqnarray}
\omega^{01}_E & = &  \frac{1}{\rho^3} \ \left[ \ \frac{ 2 M r^2 \ + \ a^2 \ [ \ M + r + ( M-r ) \cos 2 \theta \ ] }{2 \sqrt{\Delta}} \
\sigma^0_E \ + \ ra \sin \theta \sigma^3_E \ \right]\nonumber \\
\omega^{02}_E & = & \frac{a \cos \theta}{\rho^3} \ [ \ a \sin \theta \sigma^0_E \ + \ \sqrt{\Delta} \ \sigma^3_E \ ] \nonumber \\
\omega^{03}_E & = & \frac{a}{\rho^3} \ [ \ r \sin \theta \sigma^1_E \ - \ \sqrt{\Delta} \cos \theta \sigma^2_E \ ] \nonumber \\
\omega^{12}_E & = & \frac{1}{\rho^3} \ [ \ a^2 \sin \theta \cos \theta \sigma^1_E \ - \ r \sqrt{\Delta} \sigma^2_E \ ] \nonumber \\
\omega^{13}_E & = & - \frac{r}{\rho^3} \ [ \ a \sin \theta \sigma^0_E \ + \ \sqrt{\Delta} \sigma^3_E \ ] \nonumber \\
\omega^{23}_E & = &  - \frac{\cot \theta}{\rho^3} \ [ \ a \sqrt{\Delta} \sin \theta \sigma^0_E \ + \ ( r^2 - a^2 ) \sigma^3_E \ ] .
\label{36} \end{eqnarray}

We impose again that $ \sigma^3_E \ |_{r=r_{+}} \ = \ 0 $ and therefore in addition to the Euclidean time we must vary the angular variable $ \phi $

\beq \int^{\Delta t_E, \Delta \phi}_0 \ \sigma^3_E \ |_{r= r_{+}} \ = \ 0 \ \ \rightarrow \ \  \Delta \phi \ = \ \frac{a}{r^2_{+}-a^2} \ \Delta t_E . \label{37} \eeq

The condition $ \sigma^3_E \ |_{r=r_{+}} \ = \ 0 $ is necessary for two reasons:

1) to eliminate unwanted terms in the calculation of the holonomy of the spin connection;

2) in the particular case of the Kerr metric, to give a second very important constraint to the unknowns $ \Delta t_E $, $ \Delta \phi $ in addition to the banality of the spin connection, otherwise the
calculation would remain indeterminate.

We need to calculate the following intermediate contribution:

\beq \int^{\Delta t_E, \Delta \phi}_0 \ \frac{1}{2 \sqrt{\Delta}} \ \sigma^0_E \ |_{r= r_{+}} \ = \ \frac{1}{2 \rho} \ \left( \ \Delta t_E \ + \ a \sin^2 \theta \ \Delta \phi \
\right) \ = \ \frac{\rho \ \Delta t_E}{2( r^2_{+}-a^2 ) } . \label{38} \eeq

The trivial holonomy condition has as its only non-zero contribution $ \omega^{01}_E $ from which

\beq \exp \left[ \int^{\Delta t_E}_{0} \ \omega^{01}_E \ |_{r = r_{+}} \  \right] \ = 1 \label{39} \eeq

we get the final solution

\beq \Delta t_E \ = \ \frac{2 \pi ( r^2_{+} \ - \ a^2 )}{r_{+}-M} \ \ \ \ \ \ \Delta \phi \ = \ \frac{2 \pi a }{r_{+}-M}  , \label{310} \eeq

the Wick rotation of the Hawking temperature formula for the Kerr metric. We have verified that the solution (\ref{310}) coincides with the results known in the literature.
Since the Kerr metric is not diagonalizable, it is also not possible to reabsorb the period $ \Delta \phi $ into the temporal period $ \Delta t_E $, as instead we did in the case of the $BTZ$ black hole.
Maybe this property has a deep thermodynamical meaning, but we haven't been able to elaborate it.

\section{Conclusions}

To define the Wick rotation it is necessary to have a local definition of time, exploiting the isometries of the metric. This is surely possible in the case of black holes, and the Wick rotation is an natural method to calculate their thermodynamical properties. Usually this is analyzed directly by developing the Euclidean metric around the horizon of the black hole and avoiding the conical singularity that appears in the analytic continuation. In this work we have introduced an alternative method based on the vierbein and the spin connection. In this case we realize the conical singularity globally through parallel transport. To avoid
the conical singularity we have to require that the holonomy of the spin connection is trivial. This analysis confirms the necessity of a periodic Euclidean time with period related to the temperature of the black hole.
 The mechanism by which appear non trivial contributions to the spin connection is simple. The spin connection can be expressed in the basis of the vierbein with generally constant coefficients. Normally the integral of the vierbein when computing the parallel transport is null on the event horizon. Only in the case in which the coefficient in front of the vierbein component is singular at the event horizon, we obtain a non-trivial holonomy, which then gives the physical condition for calculating the period of Euclidean time.
In this article we have studied both the $ BTZ $ black hole in $ 2 + 1 $ dimensions and the Kerr metric, finding similar structures. By imposing the condition of trivial holonomy of the spin connection we found that in addition to a Euclidean time period (linked to the temperature of the black hole) there is an extra period of the angular coordinate $ \phi $. The latter is linked to the presence of the angular momentum but it is not clear whether it corresponds to another thermodynamical property of the black hole. The metric of the $ BTZ $ black hole can be diagonalized, and the calculation can be done in two different ways, but our method always gives the same answer.
In the case of the Kerr metric it is not possibile to diagonalize the metric but our calculation confirms the known results.

We believe that in the case of black holes with many degrees of freedom the spin connection analysis is simpler. Moreover this correspondence between the conical singularity of the metric and the holonomy of the spin connection can be generalized from black hole horizons to any type of horizon. In the future we could study how to derive the entropy of the black hole and its quantum corrections with the path integral method through the vierbein and the spin connection. We hope that the proposed scheme will be useful in clarifying how the Euclidean method is applied in general.

\end{document}